\documentclass[letterpaper, 10pt,conference]{IEEEtran}

\IEEEoverridecommandlockouts

\usepackage{amsmath,amssymb,amsfonts}
\usepackage{algpseudocode}
\usepackage{algorithm}
\usepackage{graphicx}
\usepackage{textcomp}
\usepackage{xcolor}
\usepackage{mathptmx}
\usepackage{siunitx}
\usepackage{float}
\usepackage{epstopdf}
\usepackage{graphicx}
\usepackage{caption}
\usepackage{subcaption}
\usepackage{comment}
\usepackage{lipsum}
\usepackage{tikz,cite}
\usepackage{pgfplots}
\pgfplotsset{compat=1.18}
\usetikzlibrary{decorations.pathreplacing,positioning}
\usepackage{caption,multirow,booktabs}
\usepackage{subcaption}
\usepackage{soul,url}
\usepackage[multiple]{footmisc}

\begin{document}

\title{UPSim: UxNB Propagation Simulator for 3D Map-Driven FR3 Deployments}

\author{\IEEEauthorblockN{Evgenii Vinogradov$^{1,2}$}
\IEEEauthorblockA{$^1$\textit{NaNoNetworking Center in Catalonia (N3Cat), Universitat Polit\`{e}cnica de Catalunya, Spain};
\\ $^2$\textit{Department of Electrical Engineering, KU Leuven, Belgium}; \\
Email: evgenii.vinogradov@upc.edu}}

\maketitle
\maketitle
\begin{abstract}
We introduce UPSim (UxNB Propagation Simulator), an open-source, Ray Tracing-calibrated solution for spatially consistent FR3 (16.95\,GHz) air-to-ground propagation modeling in Uncrewed Aerial Vehicle (UAV) networks. Instead of executing point-to-point ray launches for every ground position, UPSim derives deterministic visibility regions directly from global 3D building geometry (3D-GloBFP) via fast shadow projection. It then augments these regions with Line-of-Sight (LOS) state-specific, UAV height- and distance-dependent path loss, spatially correlated large-scale fading, and log-logistic small-scale fading parameterized by over 136 million Ray Tracing (RT) channels. Validation against reference RT data demonstrates that UPSim accurately reproduces empirical attenuation maps, outperforming the 3GPP baseline by reducing the aggregate Kolmogorov-Smirnov distance from 0.33 to 0.04. The resulting maps enable route-based analysis of channel evolution in complex urban layouts, capturing critical trajectory-level metrics such as spatial outage persistence. Consequently, UPSim provides a highly scalable, open-source middle ground between computationally expensive full RT and purely stochastic models for mobility-aware 6G aerial network planning.
\end{abstract}


\section{Introduction}
Integrating Uncrewed Aerial Vehicle-carried Base Stations (UxNBs) and upper mid-band Frequency Range 3 (FR3, 7–24\,GHz) is a promising way to achieve the ambitious 6G performance goals. Operating at these frequencies provides wide bandwidth for high-capacity Air-to-Ground (A2G) links. Deploying a UxNB at an elevated hover altitude increases Line-of-Sight (LOS) probability across a service area\,\cite{Vinogradov2018tut}, mitigating the high diffraction losses and strong effect of obstacles on mid-band radio waves\,\cite{Cui2025}. 

UxNB coverage is well investigated for sub-6\,GHz and millimeter wave bands (FR1 and FR2), however, user mobility-related issues (contiguous outage zones, handover design) require further attention: while the UxNB mostly remains stationary, ground receivers move through the underlying landscapes. FR3 propagation is dominated by sharp spatial transitions, where turning a building corner can induce abrupt power drops. Evaluating this behavior requires channel models that preserve spatial consistency in realistic environments. 

Until now, unfortunately, FR3 and A2G channels have been treated separately. FR3 propagation has been investigated in typical indoor\,\cite{Shakya2024indoor}, terrestrial\,\cite{Shakya2025outdoor,Hu2024outdoor} and vehicle-to-base station scenarios\,\cite{Aghaei2026}. The authors used measurements\,\cite{Shakya2024indoor,Shakya2025outdoor} and Ray Tracing (RT) simulations\,\cite{Hu2024outdoor,Aghaei2026}  at 16.95\,GHz as ground truth for channel characterization and modeling.

The modeling of the A2G channel attracted significant attention from academia\,\cite{Vinogradov2018tut,khawaja2019survey} and standardization bodies\,\cite{3GPP_38901,ITU}. While most of the models are fully stochastic\,\cite{3GPP_38901,ITU,Saboor2024model,Saboor2025pedestrian}, more realistic works must rely on simulations in 3D environments. Since full RT is computationally expensive, several authors resort to semideterministic models where the LOS and non-LOS (NLOS) states are defined deterministically via RT\,\cite{Colpaert2018,Colpaert2020handover,Li2021semideterministic,Saboor2023simulator} or geometric shadow projections\,\cite{Kim2023features, cho2025placement, Vinogradov2026} and then used to generate state-specific path loss. 

\paragraph{Limitations of the state of the art}
No spatially consistent A2G channel model has been reported for FR3. Apart from the lack of FR3 support, the existing A2G channel modeling paradigms face the following practical challenges:
\begin{itemize}
    \item Stochastic models (such as 3GPP\,\cite{3GPP_38901} including A2G extensions) calculate LOS probabilities using abstract distributions rather than physical 3D maps \cite{3GPP_38901}. As a result, it cannot predict the physical extent or duration of an outage along a specific street trajectory.
    \item RT simulators provide site-specific accuracy in 3D models. However, executing point-to-point ray launches across dense 2D ground grids or sweeping UxNB hover altitudes incurs significant computational overhead.
    \item 3D building models are often inaccurate: many papers use abstract environments\,\cite{Saboor2023simulator,Kim2023features, cho2025placement,Vinogradov2026} or standard databases such as OpenStreetMap (OSM) that lack physical building height tags for over 90\% of structures\,\cite{Bernard2022}. It is a critical drawback since building height is a primary predictor of UxNB performance\,\cite{Saboor2024model}.
    \item Existing Geometry-Based Shadow Projection (GBSP) approaches\,\cite{Vinogradov2026} offer a fast alternative to RT, but they currently paired with sub-6\,GHz channels and synthetic layouts, lacking FR3 support in realistic cities.
\end{itemize}

\paragraph{Beyond state-of-the-art contributions}
To bridge geometric shadow mapping with empirical FR3 wave propagation, this paper presents an open-source UxNB Propagation Simulator (UPSim), and its underlying altitude-aware A2G channel model. The primary contributions of this work are:
\begin{itemize}
    \item An open-source, 3D map-driven simulator (UPSim)\footnote{\url{https://github.com/Eugenio86/UPSim.git}} that enhances the fast GBSP engine\,\cite{Vinogradov2026} with global-scale 3D building dataset 3D-GloBFP\,\cite{Che2024ESSD} to generate spatially consistent channel maps without computationally expensive point-to-point RT.
    \item A novel altitude- and distance-aware 16.95\,GHz A2G channel model capturing path loss, spatially correlated shadow fading, and state-dependent log-logistic small-scale fading, parameterized with 136\,million RT channels.
    \item Characterization of continuous LOS/NLOS segment lengths and spatial outage persistence along ground trajectories, providing the spatial channel metrics required for ldesign of UxNBs-enabled 6G networks in FR3.
\end{itemize}

The remainder of this paper is organized as follows. Section\,\ref{sec:upsim} details the channel model structure. Section\,\ref{sec:param} describes the 16.95\,GHz RT reference dataset and parameter fitting. Section\,\ref{sec:results} presents validation results and trajectory statistics. Section\,\ref{sec:conclusion} concludes the paper.

\section{UPSim: UxNB Propagation Simulator}\label{sec:upsim}
UPSim is a 3D map-driven simulator for evaluating spatially consistent A2G propagation from UxNB to the ground users spread over realistic urban geometry. For a fixed UxNB location and a given 3D city model, it generates LOS/NLOS and channel maps on either a dense ground grid or a prescribed route. These outputs support route-level metrics such as LOS/NLOS distances and outage intervals. Since the simulator operates in the spatial domain, the same realization can later be paired with arbitrary user speeds or mobility models.

\subsection{3D-GloBFP Dataset}
In this paper, UPSim is demonstrated on the Barcelona subset of the 3D-GloBFP dataset\,\cite{Che2024ESSD}. 3D-GloBFP is a global building dataset that associates each OSM building polygon with a height estimate, thereby providing a lightweight description of the urban 3D scene that is directly usable for propagation simulation. For UxNB studies, this offers a clear advantage over relying on OSM data popular in literature\,\cite{Benzaghta2025,Lee2025}. OSM is highly valuable for footprint extraction and semantic context, but height-related tags such as \texttt{height} and \texttt{building:levels} are optional and their global availability is below 10\% \cite{Bernard2022}.  By contrast, 3D-GloBFP provides a consistent height, which is exactly what is needed to model LOS transitions and blockage regions in a reproducible way.
\subsection{3D Urban Map Representation}
The modeled scene is a planar domain $A \subset \mathbb{R}^2$ containing buildings that can obstruct the A2G link. A UxNB is described by its horizontal coordinates $\mathbf{x}^{\text{U}} = (x^{\text{U}}, y^{\text{U}})$ and altitude $h^{\text{U}}$. Each building enters the simulator through its footprint and height, i.e., as an extruded polygon.

Ground users are restricted to the accessible outdoor region $\hat{A} \subset A$ obtained after removing building interiors and deployed at $h^{\text{UE}}$ height. This enables two complementary uses of UPSim: map-wide visibility statistics over $\hat{A}$ and route-specific analysis along a UE trajectory $\mathbf{R}$\footnote{The framework accepts arbitrary route descriptions, e.g., analytical curves or discrete waypoints $\mathbf{r}_{i} = (x^{\text{UE}}, y^{\text{UE}}) \in \hat{A}$ connected by line segments.}.

\subsection{Geometry-Based LOS Estimation}\label{sec:gen_geometry}
\begin{figure}
    \centering
    \includegraphics[width=1\columnwidth]{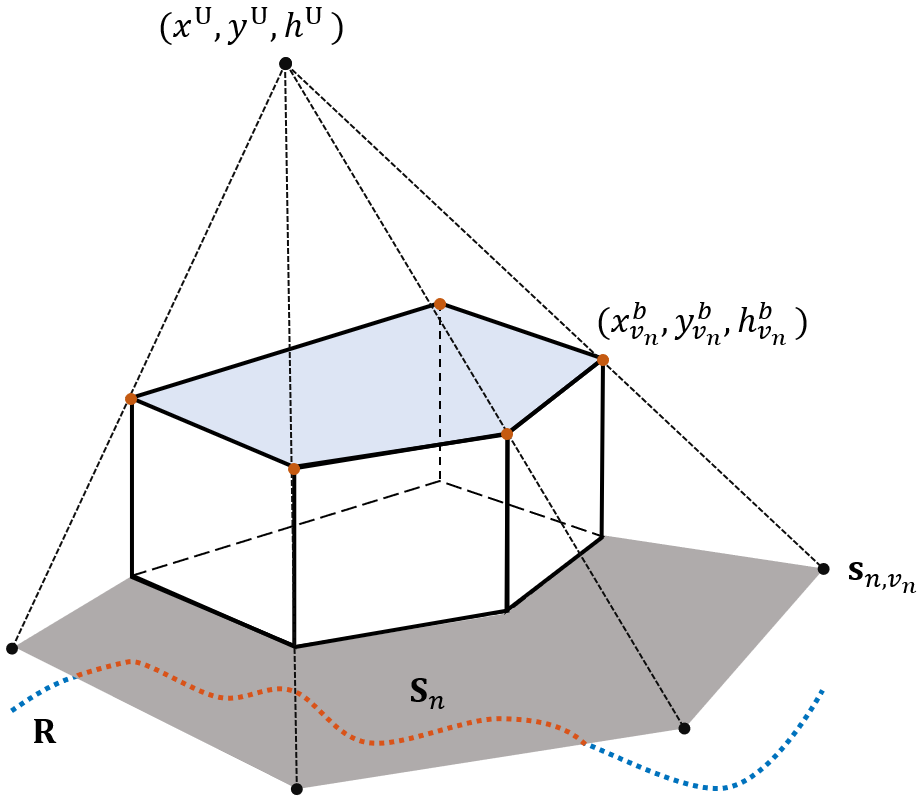}
    \caption{Illustration of the 3D building representation used in UPSim and derived from the 3D map data. The $n$-th building is described by $v_n=5$ vertices and yields the shadow region $\mathbf{S}_n$. The dotted route $\mathbf{R}$ contains LOS and NLOS segments shown in blue and red, respectively.}
    \label{fig:system}
    \vspace{-0.3cm}
\end{figure}

Let $\mathcal{B} = \{ \mathbf{b}_n \}_{n=1}^N$ denote the building set. Building $n$ is modeled as a prism with a flat roof, a polygonal footprint with $v_n$ vertices collected in $\mathbf{X}_{n}^b \in \mathbb{R}^{v_n \times 2}$, and a height $h_n^b$, i.e., $\mathbf{b}_n = (v_n, \mathbf{X}_{n}^b, h_n^b)$. The side surfaces are the $v_n$ vertical rectangles that connect adjacent footprint vertices to the roof. These primitives are sufficient to derive deterministic ground-plane blockage regions, as illustrated in Fig.\,\ref{fig:system}.

Blockage is computed by projecting each roof vertex toward the ground (User) plane with respect to the UxNB viewpoint. Assuming $h^{\text{U}} > h_{n}^b > h^{\text{UE}}$, the projected horizontal location $\mathbf{s}_{n,v_n}$ of roof vertex $v_n$ of building $n$ is
\begin{equation}\label{eq:shadow_vertex}
    \mathbf{s}_{n,v_n} = (h^{\text{U}} - h^{\text{UE}}) \frac{\mathbf{x}_{v_n}^b - \mathbf{x}^{\text{U}}}{h^{\text{U}} - h_{n}^b} + \mathbf{x}^{\text{U}}.
\end{equation}

For each building, the total blocked region $\mathbf{S}_n$ is the union of its footprint and the shadows generated by its vertical walls. The scene-wide shadow map is then obtained as
\begin{equation}\label{eq:shadow_total}
    \mathbf{S}_{\text{total}} = \bigcup_{n} \mathbf{S}_n.
\end{equation}

\paragraph{LOS map definition and LOS probability}
The set of outdoor points that remain visible from the UxNB is the complement of the total shadowed region within $\hat{A}$:
\begin{equation}\label{eq:LOS map}
    \mathbf{L}= \hat{A} - \mathbf{S}_{\text{total}}.
\end{equation}

Under a uniform user distribution over $\hat{A}$, the fraction of accessible area covered by shadows directly gives the macroscopic NLOS probability, and the LOS probability follows by complement:
\begin{equation}
P_{\text{NLOS}} = \frac{Area(\mathbf{S}_{\text{total}} \cap \hat{A})}{Area(\hat{A})},\quad P_{\text{LOS}}=1-P_{\text{NLOS}}.
\end{equation}

\paragraph{Route segmentation}
Intersecting the route with $\mathbf{L}$ and $\mathbf{S}_{\text{total}}$ yields its LOS and NLOS subsegments:
\begin{equation}\label{eq:NLOS}
    \mathbf{R}_{\text{LOS}} = \mathbf{R} \cap \mathbf{L},\quad \mathbf{R}_{\text{NLOS}} = \mathbf{R} \cap \mathbf{S}_{\text{total}} .
\end{equation}
\subsection{Channel Modeling}
For each receiver location $\mathbf{r}_{i}$, UPSim synthesizes the total channel attenuation in dB as
\begin{equation}\label{eq:total_channel}
	\Lambda(\mathbf{r}_{i}) = \Lambda_{\mathrm{PL}}(\mathbf{r}_{i}) + \xi(\mathbf{r}_{i}) + \zeta(\mathbf{r}_{i}),
\end{equation}
where $\Lambda_{\mathrm{PL}}$ is the deterministic path loss term, $\xi$ is the spatially correlated large-scale shadow fading, and $\zeta$ is the small-scale fading contribution expressed in dB. The LOS/NLOS state at $\mathbf{r}_{i}$ is taken directly from the geometry-based LOS map, while pixels inside building footprints are masked out.
\paragraph{Path loss model}
Let $d_{3\mathrm{D}}(\mathbf{r}_{i})$ denote the 3D distances between the UxNB and the UE location. The path loss used in the simulator is
\begin{equation}\label{eq:pathloss_map}
	\Lambda_{\mathrm{PL}}(\mathbf{r}_{i}) = 20\log_{10}\!\left(\frac{4\pi f}{c}\right) + 10\,n_{s}(h^{\text{U}})\log_{10}\!\left(d_{3\mathrm{D}}(\mathbf{r}_{i})\right),
\end{equation}
where $s\in\{\mathrm{LOS},\mathrm{NLOS}\}$ denotes the local propagation state. In the current implementation, the LOS Path Loss Exponent (PLE) is fixed to $n_{\mathrm{LOS}}=2$, whereas the NLOS PLE $n_{\mathrm{NLOS}}(h^{\text{U}})$ is height-dependent and can be modeled as \cite{saboor2025mmwave}
\begin{equation}\label{eq:PLE_fit}
n(h^\text{U}) = n_\infty + (n_0 - n_\infty) \exp(-h^\text{U}/h_0^n),
\end{equation}
where $n_0$ and $n_\infty$ are limiting PLEs and $h_0^n$ is a decay height.
\paragraph{Large-scale fading model}
The Large-Scale Fading (LSF, also known as shadowing) component is generated from Gaussian noise that is spatially filtered over the map, normalized, and then scaled by a LOS state and height-dependent standard deviation:
\begin{equation}\label{eq:shadow_map}
	\xi(\mathbf{r}_{i}) = \sigma_{s}(h^{\text{U}})\,S(\mathbf{r}_{i}),
\end{equation}
where $S(\mathbf{r}_{i})$ is a zero-mean, unit-variance correlated Gaussian field. 
The height-dependent standard deviation of shadow fading is modeled as in \cite{saboor2025mmwave}
\begin{equation}\label{eq:sigma_fit}
\sigma(h^\text{U}) = \sigma_\infty + (\sigma_0 - \sigma_\infty) \exp(-h^\text{U}/h_0^\sigma),
\end{equation}
with $\sigma_0$ and $\sigma_\infty$ serving as limiting values.
The filter width is controlled by the height-dependent decorrelation parameter
\begin{equation}\label{eq:ddcr_fit}
d_{dcr}(h^\text{U}) = d_{dcr, \infty} + (d_{dcr,0} - d_{dcr,\infty}) \exp(-h^\text{U}/h_0^{d_{dcr}}),
\end{equation}
with the limiting values $d_{dcr,0/\infty}$ extracted from RT data.

\paragraph{Small-scale fading model}
Finally, the i.i.d Small-Scale Fading (SSF) follows a log-logistic distribution with a LOS-state-dependent, but height-independent, shape parameter $\beta_s$. This type of fading is also often observed in aerial measurements\,\cite{Cui2019}. Note that we use the single-parameter formulation from \cite{Sanchez2022} where the scale parameter is fixed to $\alpha = \mathrm{sinc}(1/\beta)$ when the average single-to-noise ratio equals 1. The resulting fading is described by the following cumulative distribution function:
\begin{equation}
    F (\gamma) = \frac{(\gamma / \bar{\gamma})^\beta}{\mathrm{sinc}(1/\beta)^\beta +(\gamma / \bar{\gamma})^\beta}
\end{equation}
The corresponding small-scale fading term in dB is
\begin{equation}\label{eq:ssf_db}
	\zeta(\mathbf{r}_{i}) = 10\log_{10}\!\left(\gamma(\mathbf{r}_{i})\right).
\end{equation}
Consequently, UPSim produces the total attenuation map together with its path loss, large- and small-scale fading components.

\subsection{UPSim Workflow and Outputs}
\begin{figure}
    \centering
    \includegraphics[width=1\columnwidth]{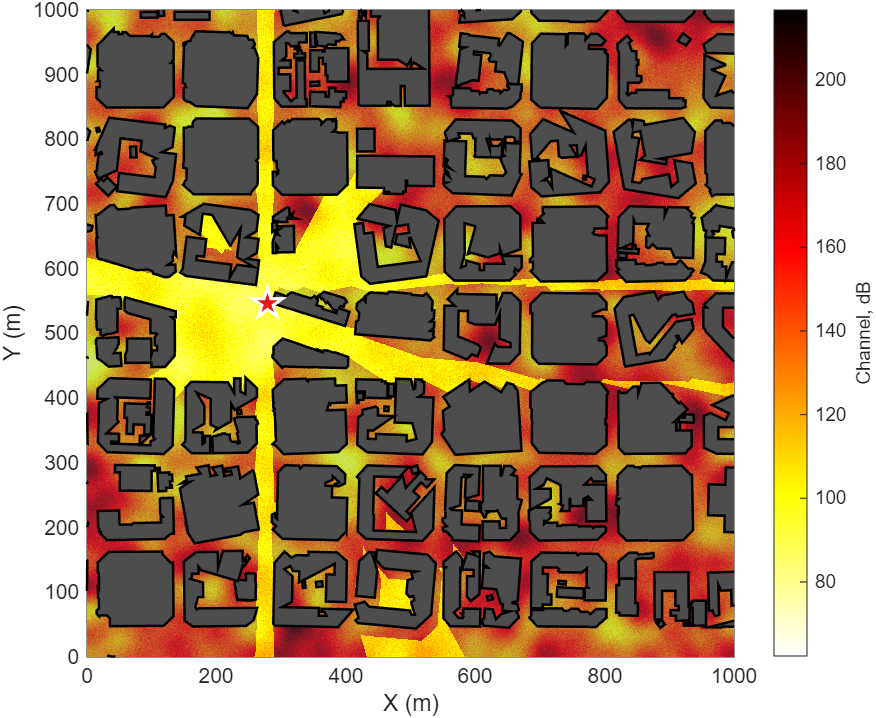}
    \caption{Illustrative UPSim output over an urban map. The UxNB is deployed 50\,m above street level; LOS and NLOS regions induced by the building-shadow geometry are evident.}
    \label{fig:simulator_ex}
        \vspace{-0.3cm}
\end{figure}

For each UxNB realization, UPSim crops the relevant buildings, constructs the visibility map through shadow projection, synthesizes the path loss and fading layers, and then derives route-level quantities such as LOS/NLOS run lengths and outage intervals. The same pipeline therefore supports both dense radio-map generation and fast evaluation along selected trajectories. Fig.\,\ref{fig:simulator_ex} shows an example of generated output.

\paragraph{Computational complexity}
RT-based LOS simulators such as~\cite{Saboor2023simulator} scale as $\mathcal{O}(N)$ and become expensive for high resolution $N$, the shadow-based approach has $\mathcal{O}(\log N)$ complexity, offering computational savings for large-scale or high-resolution route analyses. The detailed complexity derivation of the GBSP visibility engine is omitted here for brevity and can be found in our earlier work\,\cite{Vinogradov2026}. Since UPSim reuses the same geometry core, its scalability gains stem from avoiding per-location RT during LOS evaluation.

\section{FR3 RT Dataset and Parameterization of UPSim}\label{sec:param}
\subsection{FR3 RT Dataset and Setup}
The FR3 reference dataset used in this work is generated with a Cartesian ray tracing setup over a 3D city scene imported from the 3D-GloBFP dataset\,\cite{Che2024ESSD}. Building-layout information is used both to define the outdoor receiver grid and to sample valid UxNB transmitter positions. Receiver points are placed on a uniform outdoor grid, while points falling inside building footprints are excluded from the RT evaluation. Following FR3 literature\,\cite{Shakya2024indoor,Shakya2025outdoor,Aghaei2026}, we selected 16.95\,GHz as the middle of the range. The grid resolution is 20 wavelengths $\lambda$ ($\approx 0.354$\,m), which allows for separating path loss, large- and small-scale fading channel components. While this resolution only permits the extraction of first-order statistics of small-scale fading, running RT with even higher resolution is overly time-consuming. Consequently, SSF is modeled as an independent identically distributed random variable following a distribution parameterized with RT data. 

For each realization, the UE height is fixed to 1.5\,m and the UxNB is placed at a randomly generated horizontal position within the $1000\times 1000$\,m area of interest. The altitude is swept from 10 to 150\,m in 20\,m steps, and five UxNB realizations are generated per height. This yields 40 RT maps. For every map, grids of channel attenuation and LOS states are stored. The main ray tracing parameters are summarized in Table\,\ref{tab:rt_params}. 

\begin{table}
\centering
\vspace{0.1cm}
\caption{Ray tracing parameters of the FR3 reference setup}\label{tab:rt_params}
\begin{tabular}{ll}
\hline
\textbf{Parameter} & \textbf{Value} \\ \hline
Carrier Frequency & 16.95\,GHz \\
Antenna Model & isotropic \\
Total size & 1000\,m $\times$ 1000\,m \\
Simulation Resolution & $20\lambda$ ($\approx 0.354$\,m) \\
UE Height & 1.5\,m \\
UxNB Heights & 10 to 150\,m, 20\,m step \\
Tx Realizations per Height & 5 \\
Number of Reflections & 3 \\
Number of Diffractions & 1 \\
Total Number of valid channels & 136975474 \\
\hline
\end{tabular}
\vspace{-0.3cm}
\end{table}

\subsection{Channel Parameterization from RT-Derived Fits}
The current UPSim implementation is parameterized directly from RT-derived fits of the large- and small-scale channel components. Similarly to the measured channel decomposition procedure established in \cite{Saboor2026}, we use a 60-wavelength sliding window to separate SSF from large-scale channel components. We then extract the PLE per height (using all 5 Tx locations) and obtain the LSF by subtracting the deterministic path loss from the total large-scale attenuation.

Whenever a height-dependent quantity (i.e., PLE \eqref{eq:PLE_fit}, deviation \eqref{eq:sigma_fit}, and decorrelation distance \eqref{eq:ddcr_fit}) is used, we describe its behavior with a three-parameter vector $\mathbf{p}=[p_1,p_2,p_3]$ evaluated as
\begin{equation}\label{eq:exp_h_model}
	g(h;\mathbf{p}) = p_1 + (p_2-p_1)\exp\!\left(-\frac{h}{p_3}\right),
\end{equation}
which corresponds to an exponential transition toward an asymptotic value. Here, $p_1$ is the high-altitude asymptote, $p_2$ is the extrapolated low-altitude value, and $p_3$ controls the transition rate. Table\,\ref{tab:channel} summarizes the resulting parameter dependencies used by UPSim. All height-dependent parameters have higher values at the ground level. As expected, NLOS channels experience more severe LSF and SSF. Interestingly, LSF has larger decorrelation distances in the NLOS cases similarly to the measured data\,\cite{Abbas2015}.

\begin{table}[]
\vspace{0.1cm}
	\caption{FR3 Channel model parameters in UPSim}\label{tab:channel}
	\centering
	\scriptsize
\begin{tabular}{l|c|c|c}
\hline
    \textbf{Parameter} & $p_1$ & $p_2$ & $p_3$\\
    \hline
    LOS PLE, n & \multicolumn{3}{c}{2} \\
    NLOS PLE, n & 2.91 & 4.53 & 26.4\\
    LOS LSF standard deviation, $\sigma$ & 4.34 & 5.24 & 30.8 \\
    NLOS LSF standard deviation, $\sigma$ & 16.1 & 20 & 23\\
    LOS LSF decorrelation distance, $d_{dcr}$ & 7 & 14.64 & 27 \\
    NLOS LSF decorrelation distance, $d_{dcr}$ & 8.28 & 15.43 & 36 \\
    LOS SSF $\beta$ & \multicolumn{3}{c}{1.96} \\
    NLOS SSF $\beta$ & \multicolumn{3}{c}{1.91} \\
    \hline
    \end{tabular}
    \vspace{-0.3cm}
\end{table}

Accordingly, the current implementation is LOS state- and height-dependent. Height dependence enters the NLOS path loss exponent together with the LOS/NLOS shadow-fading standard deviations and decorrelation distances through Eq.\,\eqref{eq:exp_h_model}. LOS/NLOS dependence also affects the SSF statistics. However, the present SSF synthesis is not height-dependent as the data does not reveal such a dependency.

\section{UPSim Validation and Spatial Analysis}\label{sec:results}

We report two complementary evaluations using the UPSim framework. The first one validates the proposed model against 40 FR3 ray tracing maps using the same transmitter positions and carrier frequency as the ray tracing reference. It compares the empirical Cumulative Distribution Functions (CDFs) of channel attenuation obtained from RT, UPSim, and the 3GPP TR\,38.901 UMi-AV baseline\cite{3GPP_38901}. The second evaluation studies spatial effects over the Barcelona subset of 3D-GloBFP\,\cite{Che2024ESSD} by generating local LOS and channel maps around randomly placed UxNBs and by extracting trajectories from those maps.

\subsection{Channel Verification}
\begin{figure}
    \centering
    \includegraphics[width=.98\columnwidth]{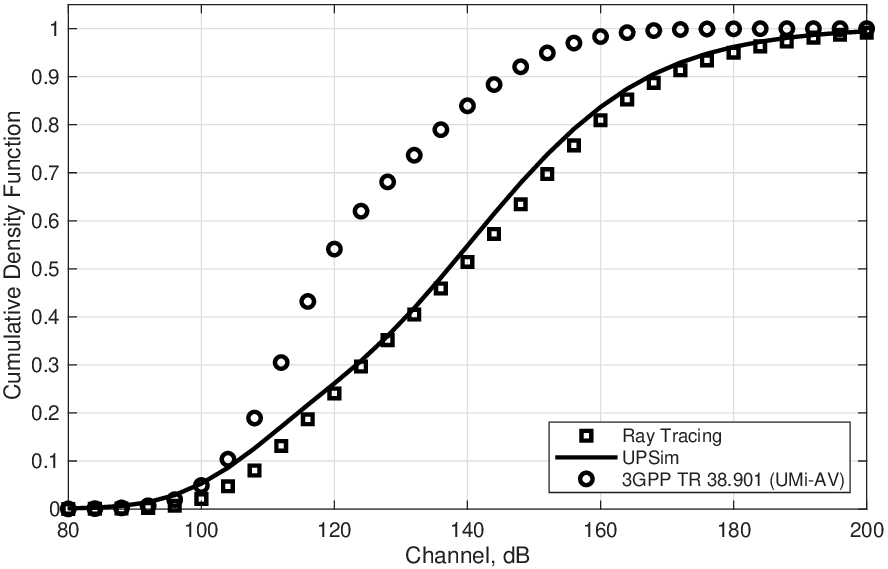}
    \includegraphics[width=.98\columnwidth]{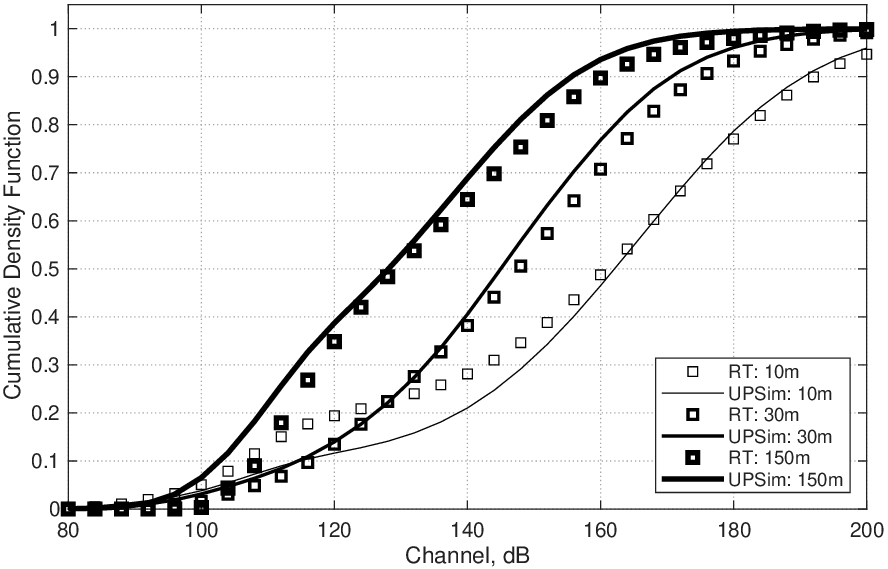}
    \caption{Validation of channel attenuation on the FR3 dataset. Top: aggregate empirical CDF comparing RT, UPSim, and the 3GPP TR\,38.901 UMi-AV baseline for the same transmitter positions. Bottom: height-specific empirical CDFs illustrating the agreement between RT and UPSim for representative UxNB altitudes.}
    \label{fig:channel}
        \vspace{-0.3cm}
\end{figure}

Fig.\,\ref{fig:channel} presents two complementary validation views for the FR3 RT dataset. The top plot shows the empirical CDF of channel attenuation aggregated over all validation maps and compares the RT reference, UPSim, and the 3GPP TR\,38.901 UMi-AV model\,\cite{3GPP_38901} evaluated at identical transmitter positions. Note that apart from the obviously inferior fit to the RT data, the 3GPP model does not provide deterministic LOS/NLOS behavior which is vital for the spatial consistency aspect.
The bottom panel of Fig.\,\ref{fig:channel} shows height-specific empirical CDFs for representative UxNB altitudes, highlighting that UPSim preserves the RT trends across different operating heights.

To quantify distributional accuracy, the Kolmogorov-Smirnov (KS) statistics ($D_{\text{KS}}$) relative to the RT reference are summarized below:
\begin{center}
\vspace{-0.1cm}
\small
\begin{tabular}{|l|c|c|c|c|}
\hline
{Model} & {10\,m} & {30\,m} & {150\,m} & {all} \\ \hline
UPSim           & 0.09 & 0.07 & 0.09 & 0.04 \\ \hline
3GPP TR\,38.901 & 0.42 & 0.36 & 0.45 & 0.33 \\ \hline
\end{tabular}
\vspace{-0.0cm}
\end{center}

\subsection{Route-Level LOS/NLOS Segment Statistics}
\begin{figure}[]
    \centering
    \includegraphics[width=1\columnwidth]{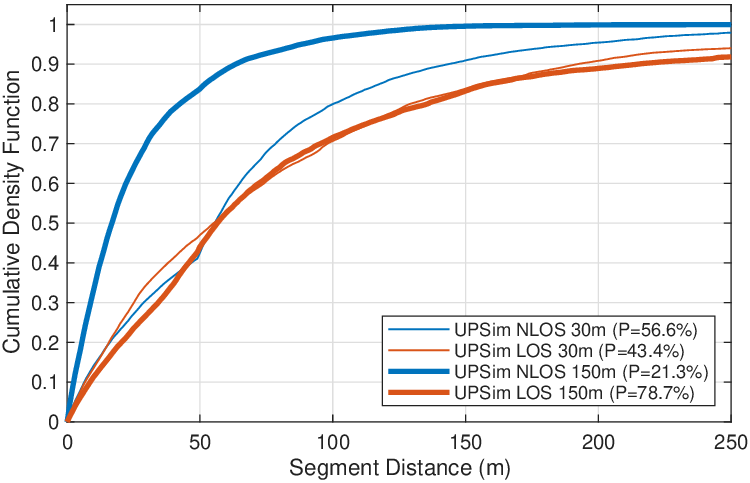}

		\caption{Empirical CDFs of LOS and NLOS segment lengths measured along trajectories extracted from UPSim maps over Barcelona. Results are reported for UxNB heights of 30 and 150\,m; the associated LOS/NLOS probabilities highlighting the benefits of high-altitude UxNB deployments.}
    \label{fig:los_cdf}
    \vspace{-0.3cm}
\end{figure}

Fig.\,\ref{fig:los_cdf} reports the empirical distributions of LOS/NLOS distances traveled along the extracted routes. Since the study is carried out on one real 3D city scene rather than on a family of parameterized synthetic environments, the curves should be interpreted as statistics for that specific Barcelona layout. For this experiment, we use a larger portion of the Barcelona map (approximately 4.5$\times$3\,km area with UxNB coverage diameter of 500 meters with 1 meter resolution). We simulate 100 UxNB locations per height. The comparison between 30 and 150\,m highlights how increasing the UxNB deployment height increases the LOS probability. Interestingly, the LOS distances remain almost identical for both UxNB deployment heights, while the NLOS zones become shorter at higher heights. Note that the CDFs of NLOS distances has "a step" around 50 meters. Similar behavior, though much more pronounced, was predicted in works using an abstract Manhattan grid\,\cite{Vinogradov2026,vinogradov2025prob} which resembles the Barcelona grid visible in Fig.\,\ref{fig:simulator_ex}.

\subsection{Threshold-based Outage Distances}
\begin{figure}
    \centering
    \includegraphics[width=1\columnwidth]{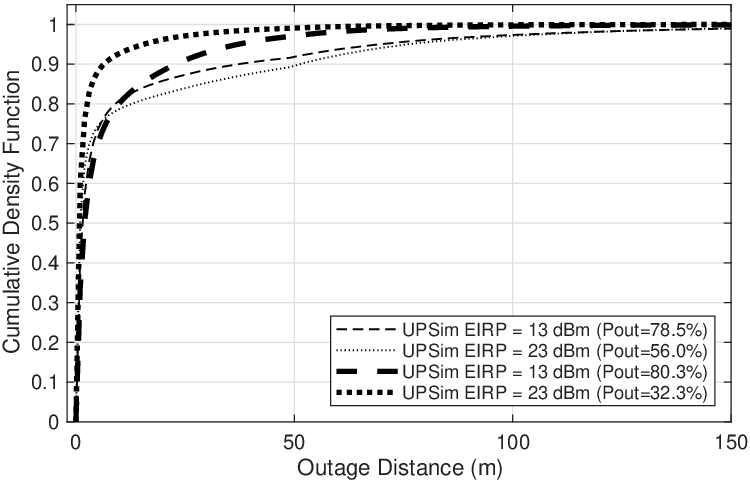}
	\caption{CDF of outage distances extracted from the Barcelona city model for UxNB heights of 30 and 150\,m. An outage is declared when the received power drops; the legend reports the corresponding fraction of traversed distance in outage.}
    \label{fig:outage}
        \vspace{-0.3cm}
\end{figure}

We define an outage segment as a contiguous trajectory interval where the channel attenuation exceeds a prescribed threshold. The UE’s receiver sensitivity is –84.7 dBm as per the 3GPP recommendation\,\cite{3gppUE}. Strict Size-Weight-and-Power (SWAP) constraints for UxNB result in moderate transmit power levels and simple antennas with low gain. In this section, we assume an equivalent isotropically radiated power (EIRP) of 13 and 23 dBm. This experiment reuses the previous setup (4.5$\times$3\,km area with UxNB coverage diameter of 500 meters, 100 UxNB locations per height). Fig.\,\ref{fig:outage} shows the empirical CDF of these outage-segment lengths for UxNB at 30 and 150\,m deployment heights. The outage probability values reported in the legend correspond to the fraction of traversed distance that is in outage under each threshold. As expected, the outage probability and experienced outage distances decrease for higher UxNB deployment heights and transmit powers. Even for 150 m, we still have around 32\% outage probability when high transmit power is used. Further investigation revealed that the outage zones are mostly concentrated around buildings. In other words, pedestrian users will have more outage. This effect was predicted theoretically in our previous study\,\cite{Saboor2025pedestrian}. Note that we again observe a step in the CDF. This indicates that the NLOS propagation is the main mechanism behind outage.

\section{Conclusions and Outlook}\label{sec:conclusion}
UPSim converts 3D city geometry and fixed UxNB positions into spatially consistent channel maps without the computational overhead of running point-to-point ray tracing at every receiver location. Integration of an altitude- and distance-dependent A2G channel model in FR3 allowed UPSim to successfully reproduce the attenuation distributions observed in RT, providing a more accurate and spatially consistent model than the standard 3GPP TR\,38.901 UMi-AV baseline. Furthermore, utilizing the Barcelona subset of 3D-GloBFP, the framework enhanced realism efficiently yields route-level statistics, such as LOS/NLOS run lengths and outage-interval persistence for varying UxNB heights. To facilitate further research and reproducibility in the community, the UPSim framework is provided as a publicly available open-source tool.
\section*{Acknowledgments}
{This work was supported by the the Spanish Ministry of Science, Innovation and Universities MICIU/AEI/10.13039/501100011033 and the European Union NextGenerationEU/PRTR through the Ramón y Cajal grant RYC2024-051003-1.}
\bibliographystyle{IEEEtran}
\bibliography{main}

\end{document}